# Determination of the Bending Rigidity of Graphene via Electrostatic Actuation of Buckled Membranes

*Niklas Lindahl[1†], Daniel Midtvedt[2†], Johannes Svensson[3], Oleg A. Nerushev, Niclas Lindvall[5], Andreas Isacsson[2] and Eleanor E. B. Campbell[4,6]\*.*

1. Dept. of Physics, University of Gothenburg, SE-41296 Göteborg, Sweden

2. Dept. of Applied Physics, Chalmers University of Technology, SE-412 96 Göteborg, Sweden

3. Electrical and Information Technology, Lund University, SE-22 100 Lund, Sweden

4. EaStCHEM, School of Chemistry, Edinburgh University, Edinburgh EH9 3JJ, Scotland

5. Department of Microtechnology and Nanoscience, Chalmers University of Technology, SE-412 96Göteborg, Sweden

6. Division of Quantum Phases and Devices, Dept. of Physics, Konkuk University, Seoul 143-701, Korea

†: These authors contributed equally to the work

\*Eleanor.Campbell@ed.ac.uk

**The small mass and atomic-scale thickness of graphene membranes make them highly suitable for nanoelectromechanical devices such as e.g. mass sensors, high frequency resonators or memory elements. Although only atomically thick, many of the mechanical properties of graphene membranes can be described by classical continuum mechanics[1,2]. An important parameter for predicting the performance and linearity of graphene nanoelectromechanical devices[3] as well as for describing ripple formation[4] and other properties such as electron scattering mechanisms[5], is the bending rigidity, $\kappa$. In spite of the importance of this parameter it has so far only been estimated indirectly for monolayer graphene from the phonon spectrum of graphite[6], estimated from AFM measurements[7,8] or predicted from *ab initio* calculations[1,9,10] or bond-order potential models[3,11]. Here, we employ a new approach to the experimental determination of $\kappa$ by exploiting the snap-through instability in pre-buckled graphene membranes. We demonstrate the reproducible fabrication of convex buckled graphene membranes by controlling the thermal stress during the fabrication procedure and show the abrupt switching from convex to concave geometry that occurs when electrostatic pressure is applied via an underlying gate electrode. The bending rigidity of bilayer graphene membranes under ambient conditions was determined to be $35.5^{+20}_{-15}$ eV. Monolayers have significantly lower $\kappa$ than bilayers.**

For deformations on a scale large compared to the inter atomic spacing, the mechanical properties of single layer graphene (SLG) as well as few layer graphene (FLG) can be modeled using the theory of two dimensional (2D) membranes. In this theory, the effective free energy is a functional of the transverse displacement $w$ and the in-plane displacement vector $u$[12]

$$F = \frac{1}{2} \int d^2x \left[ \kappa (\nabla^2 w)^2 + \mu u_{\alpha\beta}^2 + \frac{\lambda}{2} u_{\alpha\alpha}^2 \right]. \tag{1}$$

Here $u_{\alpha\beta} = (\partial_\alpha u_\beta + \partial_\beta u_\alpha + \partial_\alpha w \partial_\beta w)/2$ is the strain tensor and the indices α and β run over the Cartesian coordinates $x$ and $y$ in the plane of the graphene sheet. Repeated indices are summed over. The material parameters in (1) are the bending rigidity $\kappa$ and the Lamé coefficients $\mu$ and $\lambda$. Due to thermal fluctuations, for instance ripples, these parameters will in general depend on temperature $T$.

While the combination C ≈ ($\lambda+2\mu$) corresponding to the 2D elastic modulus has been measured at room temperature to be close to its predicted zero temperature value for graphene $C \approx 340$ Nm$^{-1}$ [1,13] a direct measurement of the bending rigidity $\kappa$ is lacking both for SLG as well as FLG. The value often quoted for the bending rigidity of monolayer graphene ($\kappa$=1.2 eV) was estimated from the phonon spectrum of graphite[6].

Using equation (1) is equivalent to treating the suspended membrane as a thin plate with a Young's modulus $E$, Poisson's ratio $\nu$, and thickness $h$, if we make the identifications

$$\frac{Eh}{1-\nu^2} = \lambda+2\mu, \quad \nu = \frac{\lambda}{\lambda+2\mu}, \quad \kappa = \frac{Eh^3}{12(1-\nu^2)}. \tag{2}$$

The parameters $Eh$, $\nu$ and $h$ are then uniquely mapped onto the parameters $\kappa$, $\mu$ and $\lambda$ of Eq. (1). Here, $E$ is not independent of $h$, rather it is the product $Eh$ which is determined. Often, as in experiments on SLG and FLG nano resonators [10,14,15], in-plane stress dominates and the first term in (1) can be disregarded. In such cases one often sets $h$=3.4 Å for SLG, the inter-planar distance between the atomic layers in graphite. As $Eh \approx 340$ Nm$^{-1}$ [1,13], this leads to $E \approx 1$ TPa. However, from Eq. (3) these values of $E$ and $h$ give $\kappa \approx 20$ eV, an order of magnitude larger than the ≈ 1 eV estimated from phonon measurements and *ab-initio* calculations.

For SLG the discrepancy stems from the different physical origins of bending rigidity in SLG and continuum thin plates. In thin plates the nonzero $\kappa$ originates from the compression/extension of the medium on either side of the neutral surface. In SLG which is not a continuum in the direction perpendicular to the membrane, bond order models have indicated two physical origins. One is due to the bond angle effect and the other results from the bond-order term associated with the dihedral angles[11]. Indeed, bond-order calculations give for SLG a $T$=0 value $\kappa \approx 1.4$ eV which is close to *ab initio* predictions of 1.46 eV [1] or 1.6 eV [9,10] and to the experimental value derived from the phonon spectrum of graphite (1.2 eV) [6]. For $T > 0$, ripples in SLG are predicted[4] to increase $\kappa$ at long wavelengths. For FLG one expects to approach the thin plate theory scaling, $\kappa \propto h^3$, as the number of layers grows. For bilayer graphene (BLG) and trilayer graphene (TLG), *ab initio* calculations and estimates using bond-order potentials have, for $T = 0$ K, predicted $\kappa_{BLG} \approx 160$ eV – 180 eV and $\kappa_{TLG} \approx$ 660 eV – 690 eV [9,16]. In these calculations, the contributions to $\kappa$ come mainly from the energy required to stretch/compress the upper/lower graphene layer as in thin plate theory. However, in contrast to SLG, where thermal fluctuations are predicted to increase $\kappa$, for FLG at $T > 0$ K, local thermal inter-plane distance fluctuations have been predicted to soften the bending rigidity [12], approaching $\kappa_{BLG} \approx 2\kappa_{SLG} \approx 3$ eV at room temperature. The large deviation between $\kappa(T = 0$ K$) \sim 10^2$ eV and the finite temperature estimate of a few eV makes it important to experimentally determine the value of $\kappa$ for $T > 0$ K.

Nano-indentation measurements have been used to extract values for the bending rigidities of thick, suspended FLG ($\geq$ 8 layers)[7]. In such experiments a force vs deflection curve is obtained by pushing the suspended part of the sample with an AFM-tip. However, extracting $\kappa$ in this way is problematic for two reasons. The first comes from the large in-plane stiffness of graphene which implies that a deformation of $w \sim 1$ Å will cause stretching contributions to dominate. Secondly, suspended samples are commonly under finite tensile strain due to electrode adhesion effects. In spite of the inherent difficulties with the technique, the extracted values fit reasonably well to the results of modelling the suspended membranes as thin plates. Using Eq. (2) to fit the data of Poot and van der Zant [7] yields (measuring $h$ in nanometres) $\kappa = 570h^3$ eV , E=0.92 TPa and $\nu = 0.16$ . A second AFM technique studies the deformation that graphene layers produce on a micro-corrugated elastic surface[8]. A model is then used to extract a so-called flattening factor that can be related to the bending rigidity as a function of the number of layers. This technique also contains uncertainty with respect to the influence of tension and interface strength. The best fit for the dependence of $\kappa$ on $h$, yields $\kappa = 182h^3$ eV (with $h$ in nanometres).

In this letter we avoid the problems inherent to the AFM measurements discussed above by exploiting snap-through instabilities in pre-buckled graphene. In the fabrication of suspended samples (beams and circular/elliptic drums) a controlled compressive strain is built in before under-etching the devices to produce the suspended SLG and FLG. When released, this leads to convex buckled geometries with zero built-in strain. The suspended regions are, in most of our samples, buckled upwards, away from the substrate. We attribute this to adhesive forces between the graphene and the electrodes. This effect of adhesion to the clamping points, which in our case is a result of under etching, has been observed also for graphene on top of trenches[13]. By biasing a backgate, an electrostatic pressure is applied to the membranes. Our method is based on relating the snap-through voltage to the local curvature, measured by AFM, and observing at what pressure the membrane undergoes a buckling deformation.

To extract $\kappa$ we note, from the analogy with thin plate theory, that our buckled membranes are expected to show similar deformation properties to those of convex shells. Fully clamped shells display buckling instability under external pressure which is observed as a snap-through from locally convex to locally concave buckling[17] at a critical pressure $p_c$. The development of a shell under external pressure is sketched in figure 1 a-c. When pressure is applied a shell with non-zero Gaussian curvature deforms first locally in the region around the structurally weakest point (see Fig. 1a). While this deformation lowers the energy due to pressure-volume (PV) work, it is at the expense of increasing the contributions from elastic energy (mainly stretching/compression). For small deformations, the balance between the elastic energy and the PV-term makes the system stable. For a deformed region larger than a critical size however, it becomes energetically favourable to form a large angle bend (Fig 1b) surrounding an inwards bulge. Following Pogorelov[17], we assume that this inward bulge forms a mirror reflection of the original surface in a plane perpendicular to the symmetry axis. Inside the bulge, the curvature of the deformed shell is then identical in magnitude to the initial surface. Hence, in this region the elastic energy density remains unchanged. The major contribution to it comes instead from a narrow region around the edge of the bulge, in the figure denoted by $\Gamma$. This energy is given by $U = \sqrt{4\kappa n(\lambda + 2\mu)} \int_\Gamma ds\, k_n^2 / k_\Gamma$, where $k_\Gamma$ is the curvature of $\Gamma$ and $k_n$ is the normal curvature of the shell along $\Gamma$. As the work done by the pressure is proportional to the area inside the bulge, this state is unstable and the edge of the bulge propagates outwards. This continues until the propagation is hindered by the edges or defects in the sample, at which point the shell is said to have "snapped through" (Fig 1c). A detailed calculation, following Pogorelov [17], for our fully clamped structures gives the following expression for the pressure at which the critical deformation is reached

$$p_c = \frac{4\sqrt{\kappa n(\lambda + 2\mu)}}{R_1 R_2} \quad . \tag{3}$$

Here $R_1$ and $R_2$ are the principal radii of curvature at the point where the instability starts and $n$ is the number of graphene layers. As $(\lambda + 2\mu) \approx 340$ Nm$^{-1}$ we can use relation (3) to extract $\kappa$ from measured values of $p_c$ and $R_{1,2}$. For the beams one can similarly show that

$$p_c \propto \frac{\sqrt{\kappa n(\lambda + 2\mu)}}{\sqrt{R_1^3 R_2}} \quad , \tag{4}$$

where $R_1$ is the curvature in the direction along the long axis of the beam.

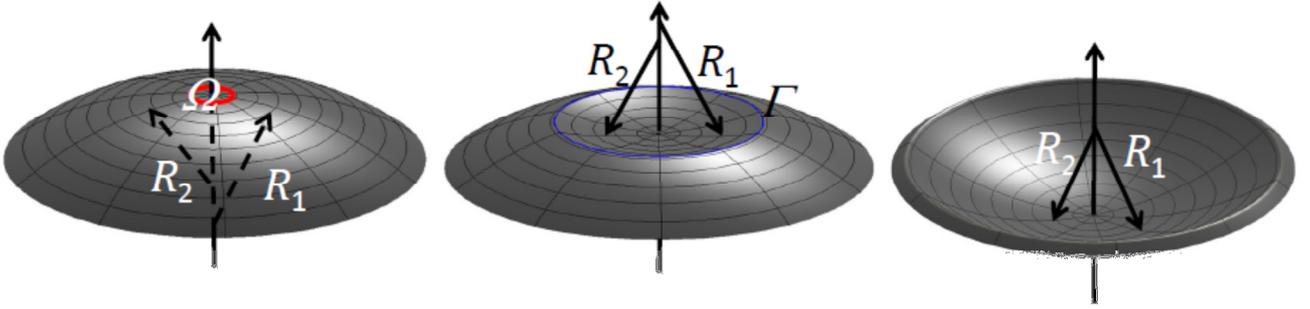

**Figure 1:** (a) – (c) Schematic pictures showing the snap-through of a convex shell. (a) For pressure smaller than a critical pressure $p_c$, a small finite deformation is formed in the region $\Omega$. (b) As the critical pressure $p_c$ is reached it becomes energetically favorable to form a concave region where the elastic energy is confined to a narrow region around the annulus $\Gamma$. (c) As the concave configuration in (b) is unstable, the deformation propagates outwards, the membrane "snaps-through". By measuring the radii of curvature $R_{1,2}$ and relating the pressure to the applied backgate voltage when the membrane snaps though, $p_c$ can be determined.

Figure 2 shows two suspended BLG beams fabricated using the techniques detailed in the Methods section. Fig. 2a shows an exaggerated schematic illustrating the way in which the beams are attached to the electrodes and the convex curvature produced when the substrate is etched away from the graphene. From the measured AFM profile in Fig. 2b it can be clearly seen that the beams are buckled to give a convex geometry. In this example, the lengths of the beams are on average 0.12 % longer than the horizontal end-to-end-distance. The buckling is also just detectable in the SEM picture in Fig. 2c where it is also possible to observe the under-etching of the electrodes.

The observed buckling is a consequence of the mismatch between the thermal expansion coefficients of the graphene and the underlying $SiO_2$ substrate, due to thermal cycling prior to etching (see Methods). Hence, before under-etching, the thermal cycling results in a compressive strain in the graphene lying on the $SiO_2$. The buckling arises upon release (etching) as the built-in compressive strain causes the sheet to have an excess length compared to the distance between the clamping points. Evidence for this is provided in the form of temperature-dependent Raman measurements detailed in the Supporting Information. The results are very reproducible in the sense that suspended membranes made from the same graphene sheet and having undergone the same thermal cycling show the same amount of built-in compressive strain before under etching and the same relative extension after under-etching. Hence, the buckling in our samples could be controlled by the extent of thermal cycling to which the substrate was exposed. In particular, for samples where thermal cycling was avoided during lithography no buckling was observed (see Supporting Information).

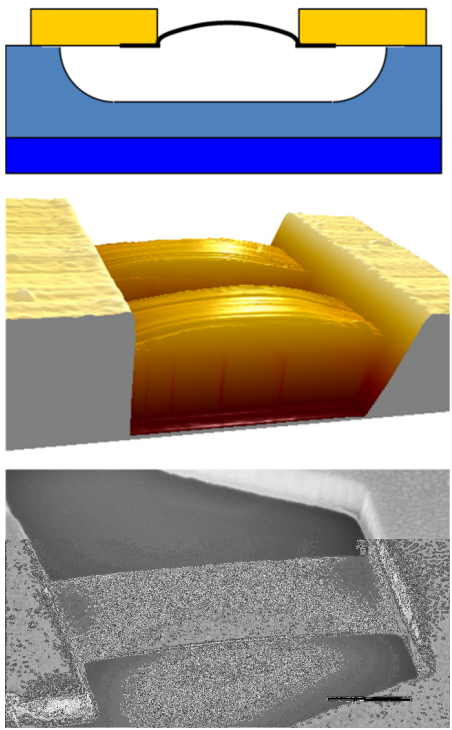

**Figure 2.** (a) Schematic picture of under-etched suspended graphene membrane. When the graphene is under-etched it is released. The built-in compressive strain together with the adhesion to the electrodes will result in a buckled shape with the graphene curving away from the substrate. (b)

AFM scan of two suspended bilayer membranes showing convex buckled shapes. (c) SEM image of a suspended membrane with visible upwards buckling. The scale bar is 1 µm.

Figure 3 shows contour plots derived from AFM scans of the beams in Fig. 2(b) that demonstrate the influence of applying a voltage to the back gate, $V_{bg}$, on the geometry of the suspended beams. Both beams clearly switch abruptly from a convex to a concave geometry as the applied voltage is increased. The switching is reversible. This is quite different behaviour to that observed for suspended graphene beams that do not have this initial buckled shape. As we have shown previously, in that case there is a continuous deflection of the suspended membrane until it snaps to contact with the underlying substrate when the pull-in voltage has been exceeded[18]. The abrupt switching can be more clearly seen in Fig. 4 where AFM line scans along the long axis of one of the bilayer beams are shown as a function of actuation voltage. In this particular example the membrane snaps from convex to concave at a voltage slightly below $V_{bg}$ = 3V. The deflection curve shown in Fig. 4d is obtained by placing the AFM tip at a fixed position in the centre of the beam and sweeping $V_{bg}$ while measuring the deflection from the initial position. This shows that there is a sharp snap-through from convex to concave buckling where the beam deflects a large distance for a small change in $V_{bg}$. For the device shown in Fig. 4, a deflection of 89 nm occurs between $V_{bg}$=2.5 V and $V_{bg}$=2.7 V. Similar switching was observed for several beams of SLG, BLG and TLG (more examples are found in the Supporting Information) with the switching becoming more abrupt and the geometry more clearly defined for BLG and TLG compared to SLG.

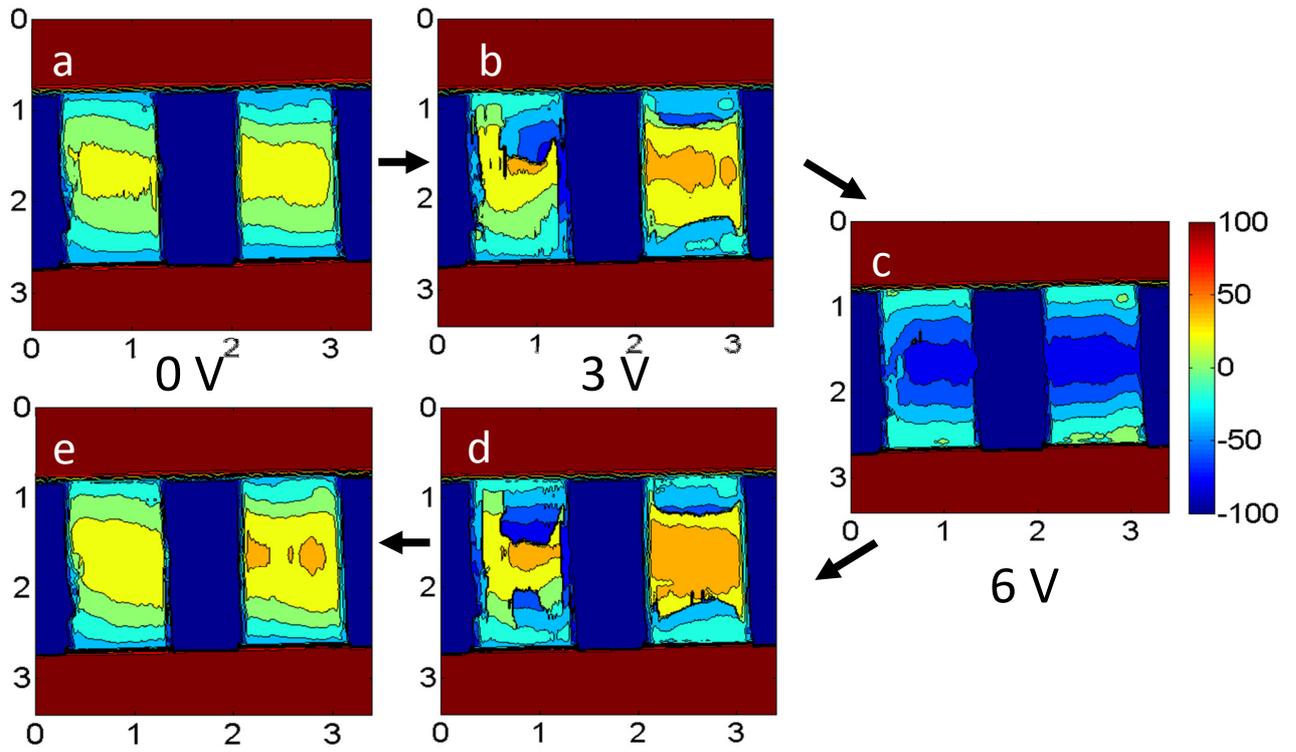

**Figure 3.** Contour plots showing the height profiles of the two suspended BLG membranes from Fig. 2b as a function of applied gate voltage, $V_{bg}$. The units of the *x* and *y*-axes are μm, the unit of the *z*-axes is nm. (a) $V_{bg}$ = 0V, (b) $V_{bg}$ = 3V, (c) $V_{bg}$ = 6V, (d) $V_{bg}$ = 3V, (e) $V_{bg}$ = 0V.

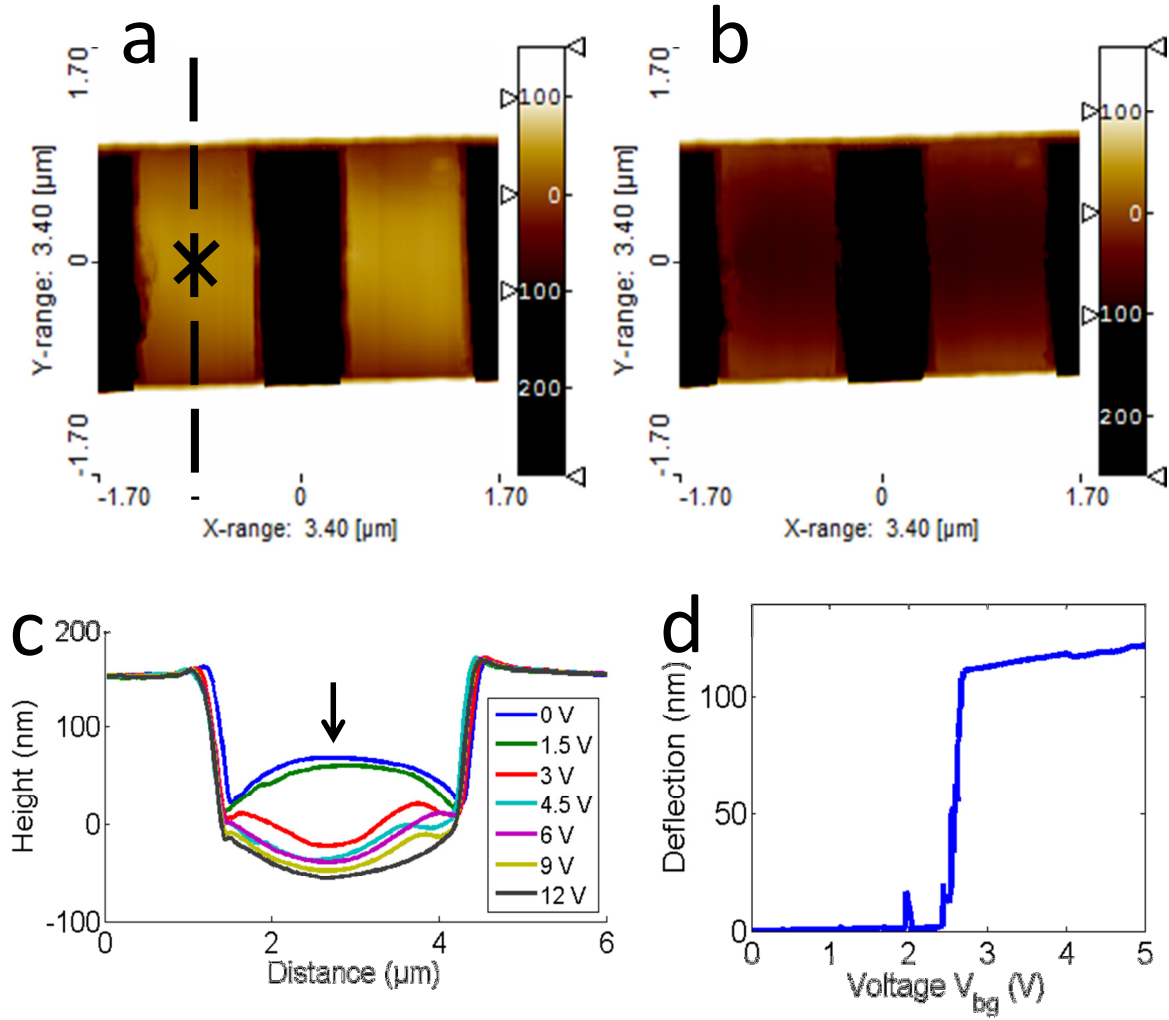

**Figure 4.** (a), (b) AFM scans of the BLG membranes from Fig.2(b) with $V_{bg}$ = 0V and 3V. (c) Line scans along the right hand membrane (following the dashed line in (a)) as a function of $V_{bg}$. (d) Deflection versus $V_{bg}$ plot for the right hand membrane measured at the position marked with a cross in (a). The deflection is defined as the deviation from the equilibrium distance at $V_{bg}$=0V measured at the centre of the beam (position indicated by arrow in 4(c)).

The beam structures clearly show a curvature along the long axis of the beam (measured between the two clamping electrodes) as is most apparent in Fig. 4(c). It is much more difficult to define a curvature across the beams, making it hard to use the observed behaviour to extract an absolute value for $\kappa$. However, we have also fabricated fully clamped membranes of circular or elliptical shape. In this case the membranes clearly show radii of curvature in two orthogonal directions and it is possible to treat them as deforming convex shells using Eq.(3) . An example of a circular suspended BLG membrane is shown in Fig. 5.

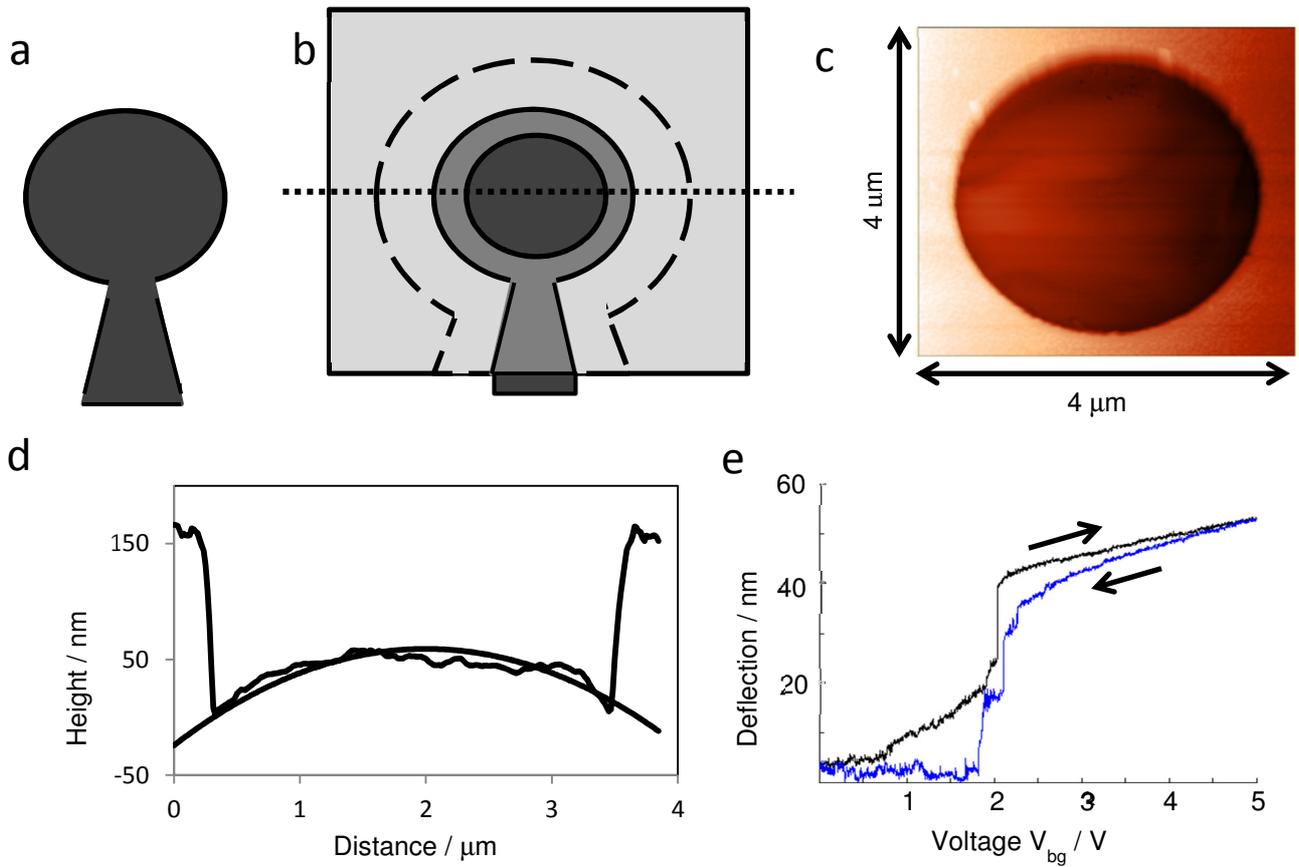

**Figure 5.** (a) Graphene "frying pan" pattern used to fabricate fully clamped circular membranes (b) A square electrode with a hole in the middle is patterned on top of the graphene, light-grey in figure. The graphene is clamped by the electrode in the light grey areas and left exposed in dark-grey areas. When the substrate is etched, the bottom-side of the handle of the "frying pan" is exposed outside the electrode. The etchant is able to penetrate freely under the graphene all the way underneath the shape of the "frying pan" and continues to under-etch the electrode, thus suspending the whole area inside the dashed line in figure 5b. (c) AFM scan of a suspended circular membrane (d) Line scan across the centre of the suspended membrane corresponding to the dashed line in (b). (e) Deflection versus $V_{bg}$ curve obtained with the AFM tip at the centre of the suspended fully-clamped BLG membrane. The deflection is defined as the deviation from the equilibrium distance at $V_{bg}=0V$ measured at the centre of the beam.

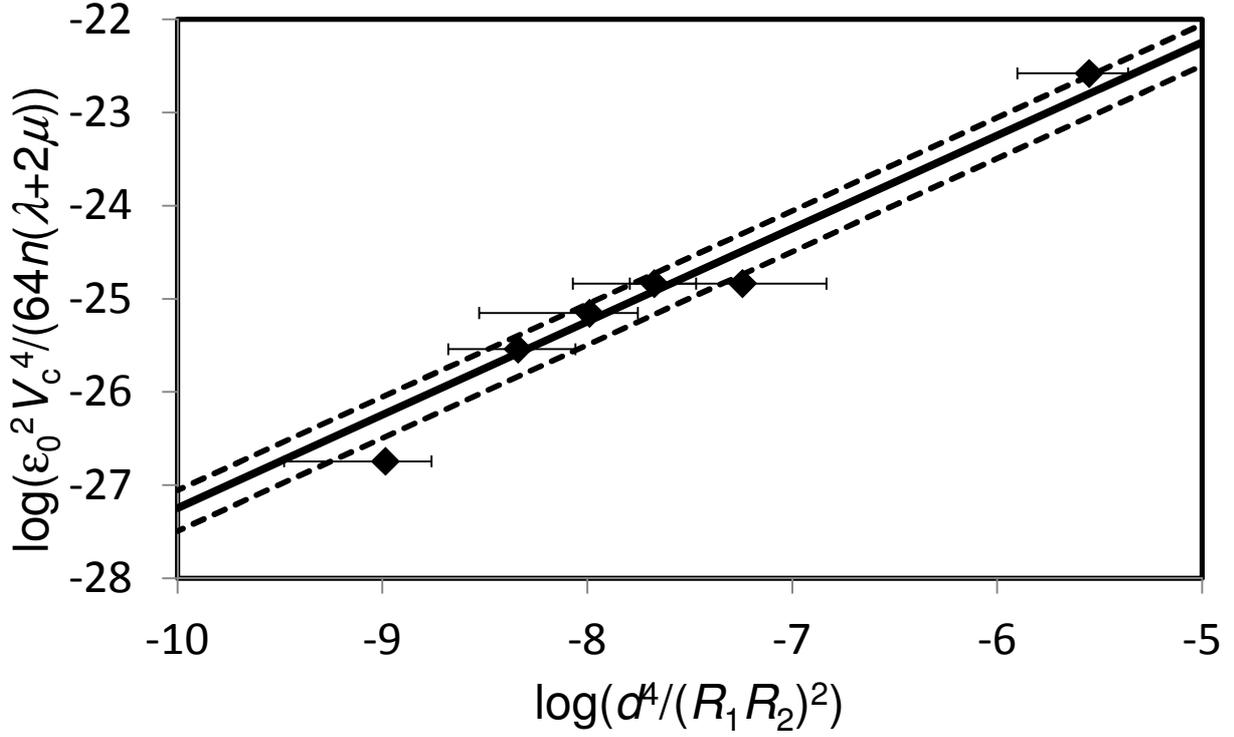

**Figure 6.** Plot used to determine $\kappa$ for fully clamped BLG. $\kappa$ is determined from the value of the y-intercept at $x = 0$. Diamonds: average experimental values obtained from fitting the radii of curvature from at least six AFM line scans on each substrate, error bars indicate the standard deviation of the fitted radii. The full line is a straight-line fit to the data and the dashed lines indicate the stated error limits.

The electrostatic pressure applied in the experiments can be calculated from the parallel plate model. Then, from Eq.(3) the bending rigidity $\kappa$ is given by

$$\kappa = \left(\frac{R_1 R_2}{d^2}\right)^2 \frac{\varepsilon_0^2 V_c^4}{64n(\lambda + 2\mu)}, \quad (5)$$

where $d$ is the effective distance to the gate (243 nm, accounting for the dielectric constant of the remaining oxide layer), $\varepsilon_0$ is the vacuum permittivity and $V_c$ the critical voltage at which snap-through occurs. The validity of the parallel plate model has been checked with FEM simulations (Supporting Information). Using $(\lambda+2\mu)=340$ Nm$^{-1}$ we plot, in figure 6, $\log\left[\varepsilon_0 V_c^4 / 64n(\lambda+2\mu)\right]$ against $\log\left[d^2/R_1 R_2\right]^2$ for the experimental devices. According to the model, the points should then fall along a straight line with unit slope. The bending rigidity can then be determined from the value of the *y*-axis intercept.

The results in Fig. 6 are for fully clamped circular and elliptical BLG membranes. We attempted to produce similar structures with SLG membranes but this proved to be very difficult and the membranes typically broke or did not show a well-defined curvature making the analysis extremely unreliable. The principal curvatures of the BLG membranes were determined by fitting the deflection data from the AFM measurements in orthogonal directions, similar to the example shown in Fig. 5(d). The results are tabulated in the Supporting Information along with the values determined for the snap-through voltage, $V_c$. The stated radii are the average values obtained from fitting at least six AFM line scans for each membrane with the error bars given by the standard deviation of the fitted radii. In order to extract $\kappa$, the gradient was constrained to be 1 (as expected from Eq. (4) and consistent with a fitted value of 1.1 ± 0.16) and the intercept was determined from a least squares fit. The fit line is shown as a full line in Fig. 6 with the estimated error limits indicated by dashed lines. The value obtained for the bending rigidity is $\kappa = 35.5^{+20}_{-15}$ eV. This value is significantly lower than the value estimated from Eq. (2), using $E$ = 0.92 TPa and $h$ = 3.4 Å, and the values obtained from zero-temperature *ab initio* calculations for BLG ($\kappa$ = 160 - 180 eV[9,16]). It is, however, considerably higher than the predicted value that assumes two independent monolayers at room temperature[16]. The agreement between continuum theory and the experimental results presented here (Eq. 5) provides convincing evidence that the continuum theory approach (Eq. 1) is valid for BLG membranes provided that one adopts a value of $\kappa$ that falls between the two extremes of the theoretical predictions.

We have also analysed the radii of curvature (along the long-axis of the beam) and critical switching voltages for a number of doubly clamped graphene beams including SLG, BLG and TLG (data included in Supplementary Information). This data has been plotted in Figure 7 in a plot of $\log(V_c^4)$ versus $\log(R_1^{-3})$, following Eq. (4). The data from the BLG doubly-clamped beams fall on a straight line in this plot. Assuming that the bending rigidity for the beams is identical to that for the BLG fully-clamped membranes, we can estimate the value of $\kappa$ for the monolayer beams by comparing the values of the y-intercept on this plot. The comparison yields an estimate of $\kappa_{SLG} = 7.1^{+4}_{-3}$ eV for the monolayer and $\kappa_{TLG} = 126^{+71}_{-53}$ eV for the tri-layer.

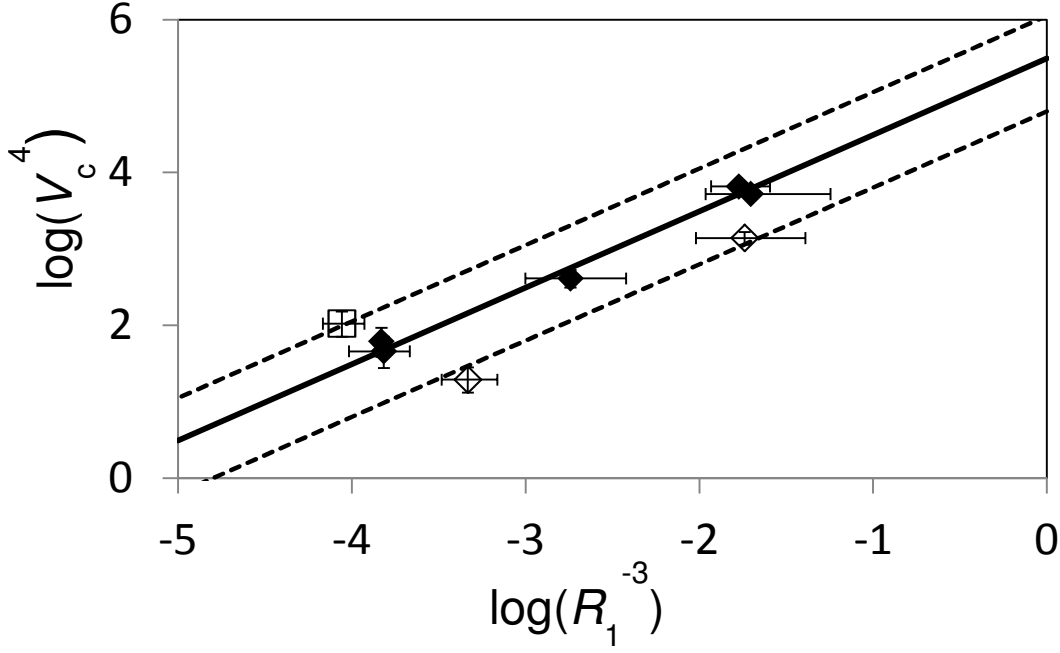

**Figure 7.** Data for doubly-clamped beams showing the expected linear behavior for a plot of $\log(V_c)$ versus $\log(R_1^{-3})$, see Eq. (4) and Eq. (5). Filled diamonds: doubly-clamped BLG; open diamonds: MLG; open square: TLG. The full line is a least squares fit to the BLG data yielding an intercept of 5.5. The dashed lines show the estimated fits for the SLG (intercept 4.8) and TLG (intercept 6.05) data. The bending rigidity can be extracted by assuming that the bending rigidity of the doubly-clamped BLG ribbons is identical to that of the fully-clamped membranes. This yields estimated values of $\kappa_{SLG} = 7.1^{+4}_{-3}$ eV and $\kappa_{TLG} = 126^{+71}_{-53}$ eV.

By studying the voltage-induced snap-through of convex buckled membranes and beams of suspended graphene we have shown that the mechanical behaviour of BLG membranes can be described within continuum theory by treating them as convex shells. The value that we obtain for the bending rigidity of BLG at room temperature ($35.5^{+20}_{-15}$ eV) is the first experimental determination of this parameter for FLG. The value lies in between the two extreme theoretical predictions for two completely independent monolayers at finite temperature and for bilayers at 0 K. An accurate experimental determination of $\kappa$ is crucial for understanding and correctly modeling the mechanical behaviour of this important new material. The method that we present here is straightforward and can easily be extended to thicker graphene layers or other thin layer materials that can be fabricated to give similar geometries.

**Methods**

Graphene was obtained from mechanical exfoliation on silicon substrates with 295 nm oxide[19]. Optical microscopy was used for finding the location of flakes with a suitable shape and number of layers. The number of layers was determined by the optical contrast and confirmed by Raman spectroscopy on selected samples. Graphene flakes were shaped into the desired geometry using electron-beam lithography (EBL) to pattern a resist mask (positive resist PMMA). The resist was typically baked at 160 °C to remove solvents after spin-coating. A low-power oxygen-plasma that removed 10 nm of resist, was used to etch the non-masked graphene. The resist mask was removed in acetone leaving the patterned graphene. A bilayer resist composed of bottom-layer copolymer MMA-MAA and top-layer PMMA was used to pattern the electrodes used to clamp and electrically contact the graphene structures. Evaporation of 3 nm Cr and 150 nm Au was done using e-gun evaporation. Cr was used as adhesion layer since it is compatible with HF-etching. A relatively thick layer of Au was used to avoid electrostatic actuation of the suspended part of the electrodes. Bi-layer resist was used to ensure an under-cut, facilitating lift-off after evaporation. Lift-off was done using ultra-sonic agitation in hot acetone. To suspend the graphene beams, the substrate was wet-etched using HF. During etching the electrodes act as an etch-mask. The etchant penetrates freely under the graphene beam. Conditions were chosen to etch away 225 nm of the underlying oxide under the entire patterned graphene structure, including the graphene covered by the electrodes. Thus to avoid excessive under-etching of electrodes, causing their electrostatic actuation during the later experiments, graphene patterns were formed first, making it possible to control the overlap distance between the electrodes and the graphene. Rinsing was done in milli-Q followed by IPA. After etching critical point drying was used to avoid collapse of the membranes due to surface tension effects during drying. Care was taken to ensure that there were no resist residues remaining on the graphene that may influence the bending rigidity measurements. It was possible to observe resist residue on supported graphene prior to substrate etching. This showed up as bright spots in the AFM height image and as dark spots in the AFM phase image. However, after etching in HF, this structure was usually removed. In order to check that any remaining resist residue did not influence the results of the bending rigidity measurements we also annealed some samples in Ar/$H_2$ and confirmed that there was no significance difference in the determined bending rigidity.

Raman spectra were obtained using a Renishaw micro-Raman spectrometer with a 514 nm excitation laser and spectral resolution better than 1 cm$^{-1}$. The shape of the 2D peak was used to confirm the number of graphene layers, estimated from the optical contrast. Raman spectra were also measured *in-situ* on the same graphene flake during heating from room-temperature to 200 °C and during cooling back to room-temperature to determine the extent of thermal stress. The results are shown in the supporting information.

Electrostatic actuation of the suspended graphene was achieved by applying a voltage, $V_{bg}$, to the silicon back-gate while keeping the graphene grounded. The depth of etching was chosen to have some remaining insulating $SiO_2$ (70 nm) to avoid a short-circuit between the graphene and the back electrode even if one or more of the graphene beams come into physical contact with the underlying substrate. Similar to previous studies of multi-walled carbon nanotubes[20] and multi-layered graphene[21], electrostatic deflection was imaged *in-situ* using AFM. The AFM was used in non-contact mode and measurements were carried out in air at 22 °C. To reduce the interaction between the suspended graphene and the AFM cantilever both were grounded. The AFM is operated under conditions where the force of interaction with the substrate is low and also operates at a frequency approximately two orders of magnitude lower than the resonant frequency of the membranes. We can therefore discount the influence of tip interactions for the substrates discussed in this paper.

**Acknowledgements**. We thank Jari Kinaret for useful technical discussions. Financial support from the Swedish Strategic Research Foundation (SSF) through "NEM arrayer för elektronik- och fotonikkomponenter" (RE07-0004) is gratefully acknowledged. AI and DM acknowledge funding from the EU 7[th] Framework Program (FP7/2007-2013) under grant agreement no. 246026, RODIN. EEBC acknowledges financial support from the WCU program of the MEST (R31-2008-000-10057-0).

**Supporting Information Available**. Raman studies of thermal cycling induced stress, residue test studies, tabulated radii of curvature and critical switching voltages.

**Contributions**

NL1, JS, NL2 fabricated the structures. NL1 carried out the majority of the experimental AFM measurements and determined the critical snap-through voltages. JS carried out initial AFM deflection measurements. OAN carried out the Raman studies. DM developed the theoretical model. NL1, DM, AI, EEBC analysed and fitted the data, interpreted the results and wrote the paper. AI supervised the theoretical work and EEBC supervised the experimental work.

**Competing financial interests**

The authors declare no competing financial interests.

**Corresponding author**

Correspondence to E.E.B. Campbell.